\documentclass[iop]{emulateapj}
\usepackage{hyperref}
\usepackage{natbib}
\usepackage{graphicx}
\usepackage{aas_macros}

\newcommand{\Lund}{\ensuremath{\mathrm{S}}}

\begin{document}
\title{Self-generated turbulence in magnetic reconnection}
\author{Jeffrey S. Oishi}
\email{jeffrey.oishi@farmingdale.edu}
\affiliation{Department of Physics, Farmingdale State College, Farmingdale, NY 11735}
\altaffiliation{also Department of Astrophysics, American Museum of
  Natural History, New York, NY} 
\author{Mordecai-Mark Mac Low}
\affiliation{Department of Astrophysics, American Museum of Natural
  History, New York, NY} 
\email{mordecai@amnh.org}
\author{David C. Collins}
\affiliation{Department of Physics, Florida State University,
  Tallahassee, FL}
\author{Moeko Tamura}
\affiliation{Department of Physics, Barnard College, New York, NY}

\begin{abstract}
  Classical Sweet-Parker models of reconnection predict that
  reconnection rates depend inversely on the resistivity, usually
  parameterized using the dimensionless Lundquist number ($\Lund$).
  We describe magnetohydrodynamic (MHD) simulations using a static,
  nested grid that show the development of a three-dimensional
  instability in the plane of a current sheet between reversing field
  lines without a guide field. The instability leads to rapid
  reconnection of magnetic field lines at a rate independent of
  $\Lund$ over at least the range $3.2\times 10^3 \lesssim \Lund
  \lesssim 3.2 \times 10^5$ resolved by the simulations. We find that
  this instability occurs even for cases with $\Lund \lesssim 10^4$
  that in our models appear stable to the recently described,
  two-dimensional, plasmoid instability. Our results suggest that
  three-dimensional, MHD processes alone produce fast (resistivity
  independent) reconnection without recourse to kinetic effects or
  external turbulence. The unstable reconnection layers
%mm are consistent with the picture of
%  turbulent reconnection advanced by Lazarian, Vishniac, and
%  collaborators.
  provide a self-consistent environment in which the extensively
  studied
%mm [> 15 years of work at this point...]
 turbulent reconnection process can occur.
\end{abstract}
\maketitle

\section{Introduction}
\label{sec:intro}
During magnetic reconnection, magnetic field lines change topology,
resulting in the conversion of magnetic energy into both thermal
energy and kinetic energy of bulk flows and non-thermal particles.
The rate at which this process occurs in the classical Sweet-Parker
picture \citep{Sweet1958,Parker1957} depends on the Lundquist or
magnetic Reynolds number $\Lund = v_A L/\eta$, where $v_A$ is the
Alfv\'en speed, $L$ a characteristic length of the system, and $\eta$
the resistivity.  The Sweet-Parker rate is orders of magnitude too
slow to explain the fast reconnection seen in low resistivity plasmas
during solar flares and sawtooth crashes in tokamaks
\citep{Yamada2010}. Because it is a fundamental plasma process,
reconnection is thought to be important in astrophysical environments
as diverse as the heliosphere \citep[e.g.][]{2010ApJ...718...72E} and
microquasars \citep{2015MNRAS.449...34K}. 

The identification 
of the 2D plasmoid instability
1\citep{1986PhFl...29.1520B,Loureiro2007,Huang2013}, a
super-Alfv\'enic, small-scale instability, has provided a mechanism to
greatly speed up Sweet-Parker reconnection. However, this instability
has primarily been studied in two dimensions (2D), assuming symmetry
in the plane of the current sheet. The reason for this dimensional
reduction is that the reconnection process is inherently multi-scale,
with a large separation between the global scale of the reconnection
layer and the resistive length where the instability grows. Even 2D
simulations tax state of the art computational resources if uniform
grids are used.

\citet{Lazarian1999} argued that reconnection in the presence of any
sort of turbulence would be fast, because the turbulence would drive
many points of contact between the opposed field lines. This idea has
been put on a rigorous mathematical basis \citep{Eyink2011} as
reviewed by \citet{2015arXiv150201396L} and
\citet{2015ASSL..407..311L}. Indeed, recent modelling suggests that
turbulent reconnection may be responsible for the radio and gamma-ray
emission from accreting black holes
\citep{2015ApJ...799L..20S}. However, these ideas all require that
reconnection proceed at a large fraction of $v_A$. Numerical models
examining reconnection in forced turbulence support this theory,
starting with \citep{Kowal2009}. In this work, we demonstrate that
turbulent reconnection proceeds in a very similar fashion when the
turbulence is self-generated from an instability of the reconnection
layer itself.

Here, we describe a set of nested grid simulations that model the
reconnection layer in 3D over a broad range of $\Lund$, without any
forcing or guide field.  These simulations show that a startlingly fast, 3D,
instability occurs in the plane of the current sheet, which was
assumed uniform in the 2D simulations. This instability drives
a large increase in the rate of reconnection, that we show remains independent
of $\Lund$ over two orders of magnitude of variation in the
resistivity.

\citet{2012PhPl...19k2901B,2013PhPl...20c2903B} has argued
that reconnection requires a point geometry to proceed, so that it is an
inherently three-dimensional (3D) process. The work we describe here demonstrates that
such a 3D geometry naturally arises even from 2D initial conditions,
resulting in fast reconnection apparently independent of $\Lund$.

Previous work in this field has shown 3D instability, but has not
provided a clear demonstration of independence of reconnection rate
from $\Lund$. \citet{2003AdSpR..32.1029D, 2005ApJ...622.1191D} focused
on the case of a current sheet with a strong guide field, and found a
3D instability set in for a weak enough guide field, which they called
a secondary instability. However, they did not measure the scaling of
the reconnection rate with $\Lund$. \citet{Lapenta2011}
reported the breakdown of an initially 2D Harris sheet into a fully 3D
reconnection region with greatly enhanced reconnection rate.
An MHD kink instability on a central plasmoid was followed by a
Rayleigh-Taylor instability driven by the reconnection jet interacting
with the plasmoids at the ends of the layer. However, again, no test
of the dependence on $\Lund$ was performed. Another numerical
experiment has shown that thin, 3D, current layers are unstable to
infinitesimal perturbations and reconnect at a rate apparently
independent of Lundquist number $\Lund$ \citep{2013arXiv1301.7424B},
but only a factor of three variation in $\Lund$ was
explored. \citet{2010ApJ...718...72E} studied the formation of coronal
current sheets due to photospheric forcing in a global, 3D, AMR
simulation. They concluded that the dynamics of the current sheet were
3D, allowing a steady rather than the bursty reconnection rate found
by 2D models of the plasmoid instability. Finally, 3D reconnection in
the collisionless limit has been explored by \citet{Daughton2011a} and
\citet{2013PhPl...20h0703P}. That work largely focused on particular
kinetic effects that drive dissipation at the smallest scales.

In Sect.~\ref{sec:methods} we describe our computational methods,
while in Sect.~\ref{sec:field} and~\ref{sec:scaling} we present our
results.  Finally, we discuss the implications in Sect.~\ref{sec:discussion}.

\section{Methods}
\label{sec:methods}
We use the mesh refinement code {\em Enzo}, which solves the
compressible, adiabatic, 
%mm [does Bryan et al. describe the resistivity module?]
         resistive, 
MHD equations \citep{2014ApJS..211...19B}. We use static refinement to focus
computational effort on the current sheets.  Our computations are
performed within a cubic, 3D volume, with periodic boundary conditions
in all three dimensions.  From the available algorithmic options, we
choose piecewise linear reconstruction, the HLLD Riemann
shock-capturing solver, and constrained transport \citep{Gardiner2005} to
ensure $\mathrm{\nabla \cdot B} = 0$ to machine precision
\citep{Collins2010}. We performed all analysis using the \texttt{yt}
toolkit \citep{Turk2011}.

Our initial condition is a pair of oppositely directed, parallel
current sheets to accommodate the periodic boundary conditions, each
perturbed following the GEM Reconnection Challenge \citep{Birn2001} to
initiate Sweet-Parker reconnection. All but two of our runs are
initialized with low-amplitude, 3D velocity perturbations with mean
Alfv\`en Mach number $\left< v/v_A \right> \sim 4.3 \times
10^{-5}$. These perturbations have a spectrum $v_k \propto k^{-4}$
with wavenumbers ranging from $k_{min} \leq k \leq k_{max}$. We choose
$k_{min}/2\pi = 10$ and $k_{max}/2\pi = 15$, except  for run C+, which has $k_{min}/2\pi = 30$ and
$k_{max}/2\pi = 35$. We do
not continue to force the velocity field during the simulation. We normalize all
lengths to the size of the box $L = 1$, densities to the initial
density at the center of the sheets $\rho_0 = 1$, and times to
the Alfv\'en crossing time of each sheet $t_{A} = \delta_0/v_A$ where
$v_A = B_0 / \sqrt{4 \pi \rho_\infty} \simeq 3.2$ is the Alfv\'en
speed of the upstream plasma. $\delta_0 = 0.02$ is the scale length of
the initial current sheet. Table~\ref{tab:run_data} lists the
parameters of our runs.

A minimum resolution requirement for reconnection is the proper
resolution of the Sweet-Parker current layer, whose width $\delta_{SP}
\simeq L/\sqrt{S}$, where $L$ is the length of the layer\footnote{A
  popular alternative is to use $\delta$ to represent the
  \emph{half}-width of the current sheet, but in that case, $L$ is the
  half-length as well.}.  We define two grid refinement regions
covering the entire plane of the current sheet ($0 < [x_{r}, y_{r}] <
1$) with a height $z_{r} \sim 12.5 \delta_0$ centered on each of the
initial current sheets. These refined regions have two levels of
refinement atop a $128^3$ base grid, leading to an effective
resolution of $512^3$ in the current sheet centers, except for run
A*, which has three levels of refinement, for an effective $1024^3$
resolution.

\begin{figure}
  \centering
  \includegraphics[width=\columnwidth]{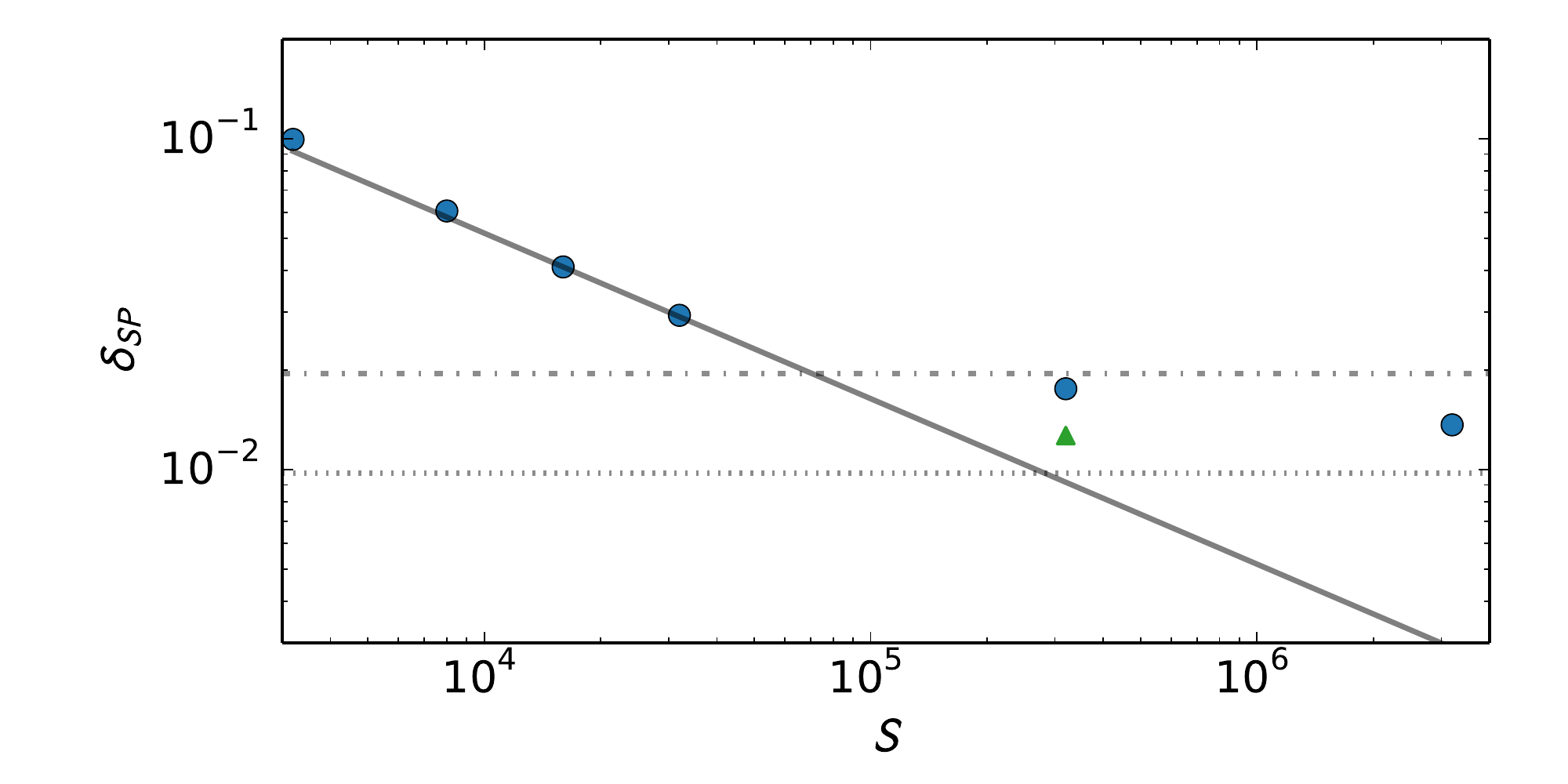}
  \caption{ Width $\delta_{SP}$ in units of box size of the current
    sheet during quiescent reconnection at $t = 75 t_{A}$, prior to
    the onset of instability. The circles show simulations with
    varying Lundquist number $\Lund$, the solid line gives the
    Sweet-Parker scaling, and the triangle shows run A* at double
    resolution. The dot-dashed line shows a resolution of 10 zones for standard
    ($512^3$-equivalent) runs, while the dotted line shows 10 zones for our
    high resolution ($1024^3$-equivalent) run.}
  \label{fig:sheet_width}
\end{figure}

Figure~\ref{fig:sheet_width} shows $\delta_{SP}$ of the initial
Sweet-Parker current sheet at $t = 75 t_{A}$, long before any
unstable perturbations have grown to significant amplitudes. All runs
with $S < 10^{5}$ have current sheet widths
that agree well with the Sweet-Parker prediction, because
they are resolved by $ \gg 10$ zones across
the sheets. The run with $S = 3.2 \times 10^5$ demonstrates the
effects of marginal resolution, while the run with $S = 3.2 \times
10^{6}$ is only resolved by $\sim 3$ zones, and is a factor of four
too thick. We do not use this last run (run J) in our subsequent analysis,
although it serves as an important limit on the numerical resistivity
of our code.

\begin{table}[htb]
\caption{Run data}
\label{tab:run_data}
\begin{tabular}{lllll}
run & $\Lund^1$         & $\eta^2$  & $\gamma^3$            & notes\\
A   & $3.2 \times 10^5$ & $10^{-5}$ & $-3.2 \times 10^{-3}$ & \\
A*  & $3.2 \times 10^5$ & $10^{-5}$ & $-3.3 \times 10^{-3}$ & double resolution\\
B   & $3.2 \times 10^4$ & $10^{-4}$ & $-5.6 \times 10^{-3}$ & \\
C   & $1.6 \times 10^4$ & $2 \times 10^{-4}$ & $-4.8 \times 10^{-3}$ & \\
C+   & $1.6 \times 10^4$ & $2 \times 10^{-4}$ & $-1.8 \times 10^{-3}$ & $k_{min}/2\pi = 30$ perturbation\\
D   & $8.0 \times 10^3$ & $4 \times 10^{-4}$ & $-2.1 \times 10^{-3}$ & \\
E   & $3.2 \times 10^3$ & $10^{-3}$ & $-1.4 \times 10^{-3}$ & \\
F   & $3.2 \times 10^2$ & $10^{-2}$ & --                    & stable to 3D instability \\
G   & $3.2 \times 10^3$ & $10^{-3}$ & --                    & no initial perturbations\\
H   & $3.2 \times 10^5$ & $10^{-5}$ & --                    & no
initial perturbations\\
J   & $3.2 \times 10^6$ & $10^{-6}$ & --                    &
underresolved, unanalysed \\
\end{tabular}
$^1$ Lundquist number
$^2$ Resistivity in code units
$^3$ Decay rate of magnetic energy (see text) 
\end{table}

\section{Field dynamics}
\label{sec:field}

\begin{figure}
  \centering
  \includegraphics[width=\columnwidth]{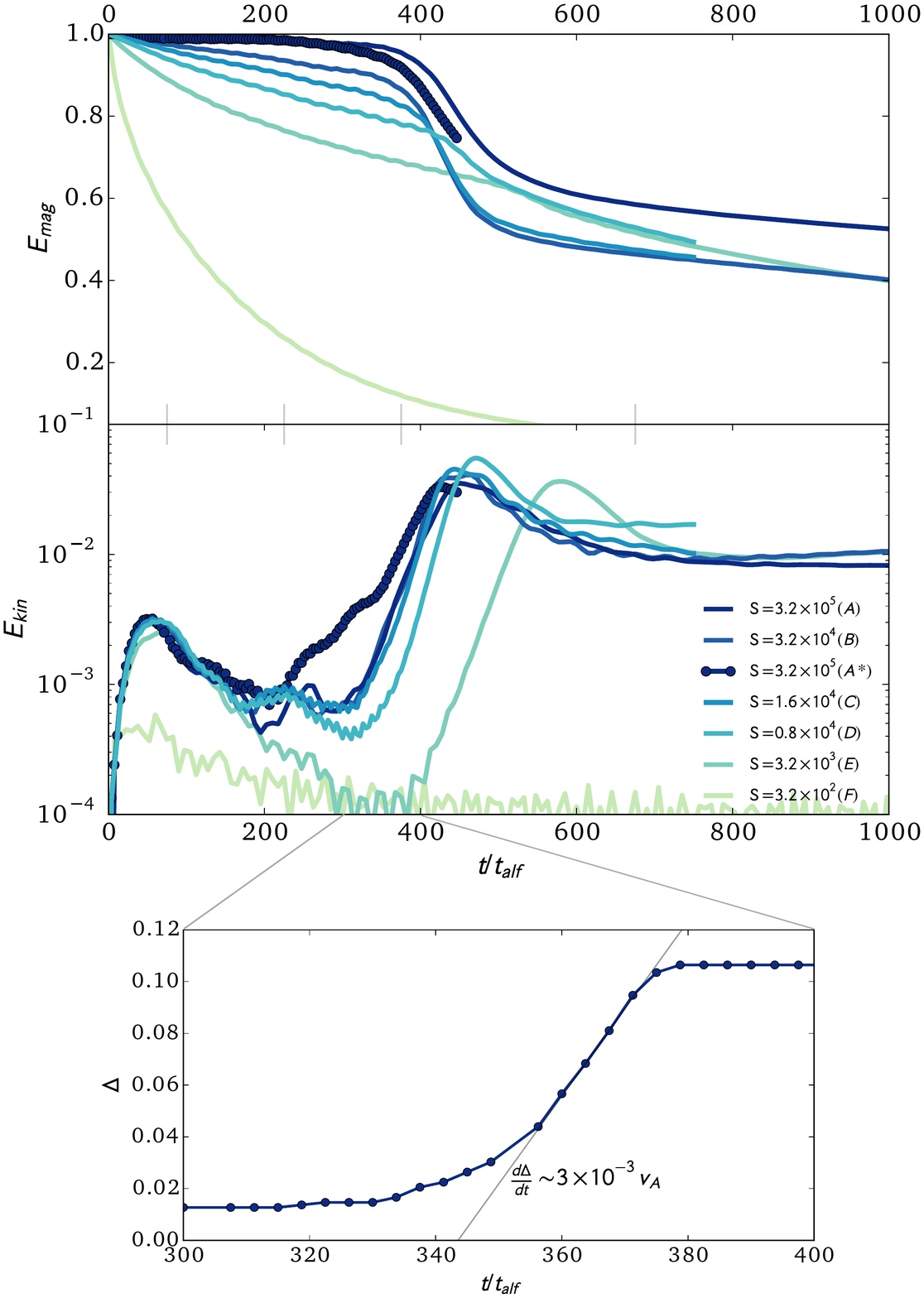}
  \caption{(upper) Volume integrated
     kinetic and magnetic energies in the
    simulation domain as a function of time for several values of 
%mm[for folks who just read the figure caption] 
    Lundquist number $\Lund$. Letters in the legend give the run name
    in Table~\ref{tab:run_data}, and the gray tickmarks on the central
    axis give the times of the four panels in
    Figure~\ref{fig:current}. (lower) Current sheet width $\Delta$ as
    a function of time for run $A*$. The light line shows a linear fit
    to the unstable period, giving a layer growth rate of $d\Delta/dt
    \sim 3 \times 10^{-3} v_A$.}
  \label{fig:energy}
\end{figure}

\begin{figure*}[htb]
  \centering
  \includegraphics[width=\textwidth]{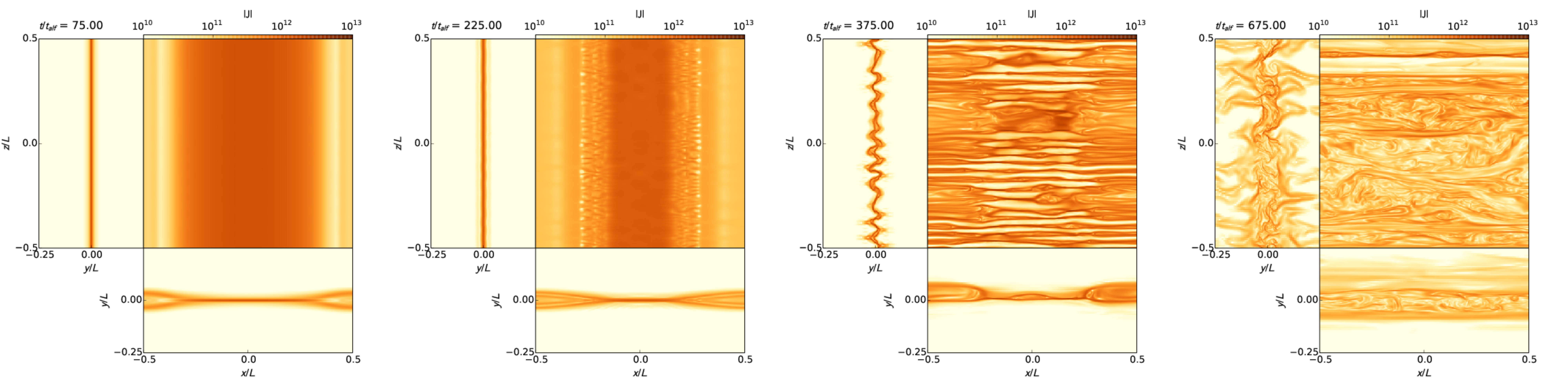}
%% this figure made with:
%% montage DD0010_current.pdf DD0050_current.pdf DD0090_current.pdf -tile 3x1 -mode concatenate output.pdf

  \caption{Slices of current density $|\mathbf{J}|$ at four times in
    run A, with $\Lund = 3.2 \times 10^5$. Each panel shows the lower
    current sheet in our simulation box (the upper sheet looks
    morphologically similar at each time). Note that the instability grows in both
    the $y$--$z$ and $x$--$z$ planes at roughly the same time. }
   \label{fig:current}
\end{figure*}

Reconnection in our models begins at the Sweet-Parker rate
 expected
for a stable field reversal, as shown by the width of the current
sheet.  This leads to the initial slow decline of
the volume integrated magnetic energy for all simulations
(Figure~\ref{fig:energy}), as well as the low values of integrated
kinetic energy.  Instability along the plane of the current sheet then
sets in, driving far faster reconnection, and transferring energy from the
magnetic field into the flow, as shown by the sudden drop in magnetic
energy and the corresponding rise in kinetic energy. The morphology of
the onset and growth of the instability is shown in the middle panels
of Figure~\ref{fig:current}, while the final panel shows its saturated
state.

The 
development of the 3D instability results in the buckling of the current sheet in the
$y$--$z$ plane, with a characteristic wavenumber $k_z/2\pi \sim 12$ (third
panel of Fig.~\ref{fig:current}). The
simulations of \citet{Lapenta2011} can be seen to show similar
behavior, though it is not emphasized in their paper.  They used a
thin box in the third dimension, so they only had two wavelengths in
that direction. %across their box.  
Ours is a factor of six deeper in the $z$ direction than
theirs\footnote{Note that the \citet{Lapenta2011} box is oriented so their $y$
  axis corresponds to our $z$ axis}. Thus, our results give a
wavenumber consistent with that shown in their figures.

We find
that resistivity stabilizes the 3D instability for $\Lund \lesssim 10^{3}$. Run D
with $\Lund = 3.2 \times 10^3$ shows the instability clearly through
the growth of kinetic energy, although the total amount of reconnection
driven by the turbulence is small, because the large resistivity has already allowed significant
laminar reconnection to occur. 

When $\Lund > 10^4$, the transition from laminar to
turbulent reconnection begins with the rapid growth of kinetic energy
%mm
    apparently 
driven by a kink-type instability along the plasmoids in the
$z$-direction. 
%mm [try to summarize JSO musings on this subject]
    However, this is not a classical kink instability, as the interior
    of the plasmoids is a demagnetized reconnection region, rather
    than a column of current-carrying plasma surrounded by vacuum as
    would be true in the classical case.  The growth of kinetic energy
    and decay of magnetic energy must be due to reconnection occurring
    where field lines are driven together by these instabilities
    rather than a simple rearrangement of the horizontal field by
    them.

It appears from our models that the 3D instability may actually grow
independently of the 2D plasmoid instability. Run E with $\Lund
\lesssim 10^{4}$ lacks evidence for the growth of the 2D 
plasmoid instability seen at higher Lundquist numbers by ourselves and previous authors
\citep{Loureiro2007,Samtaney2009,Uzdensky2010,Huang2010,Loureiro2012,Huang2013},
but nonetheless shows the growth of the 3D 
instability, albeit with a delay in 
onset of rapid growth (Fig.~\ref{fig:energy}).  

This delay
occurs because growth of the 2D plasmoid instability triggers
secondary Richtmyer-Meshkov instability \citep{Richtmyer1960,Meshkov1969}
that accelerates onset of the 3D instability, but is not
required for 3D instability to occur.  
The acceleration of a flow across the density contrast
between the plasmoid and the surrounding current sheet along the $x$
axis drives the Richtmyer-Meshkov instability (analogous to the
Rayleigh-Taylor instability that occurs in a gravitational field). This 
instability begins 
when weak transient shocks from the formation of the
plasmoids at the sides of the domain ($x = -0.5, 0.5$) pass over the
density gradient produced by the first plasmoid to form around the initial X
line at $(x, y) = (0, 0.5)$. The secondary instability drives weak, initial mixing along
the X line.

We examined three numerical issues with further runs. First, to
determine if resolution affects our major result, we run our marginally resolved run A at
twice the resolution (run A*). This run results in essentially identical growth rate
$\gamma$. 

Second, we checked that the instability is not purely numerical
by performing Runs G \& H, identical to the unstable Runs E \& B
respectively, except without any initial perturbations. The
instability did not grow in either of these runs. In Run G, which has
$\Lund = 3.2 \times 10^3$ and is thus stable to the 2D plasmoid
instability, the entire 3D volume settled down into a steady, laminar
reconnection at the Sweet-Parker rate, with a sheet width $\delta
\simeq \delta_{SP} = 1.72 \times 10^{-2} L$. Run H, on the other hand,
is unperturbed in the $y$--$z$ plane, but is unstable to the plasmoid
instability. In this case, we find a vigorous plasmoid instability
that is entirely symmetric along the $z$ axis, demonstrating that our
code is sufficiently stable to recover the 2D results in 3D if there
are no explicit 3D perturbations. 

Third, run C+ was performed at
standard resolution with smaller scale perturbations ($k_{min}/2\pi =
30$). We find the growth rate is higher by a factor $\sim 3$ for the
lower $k$ perturbations, suggesting a wavenumber dependence for the
underlying instability. We will pursue a formal stability analysis of
the instability in a separate paper, and this wavenumber dependence
represents an important test for that work.

\section{Scaling}
\label{sec:scaling}

\begin{figure}
  \centering
  \includegraphics[width=\columnwidth]{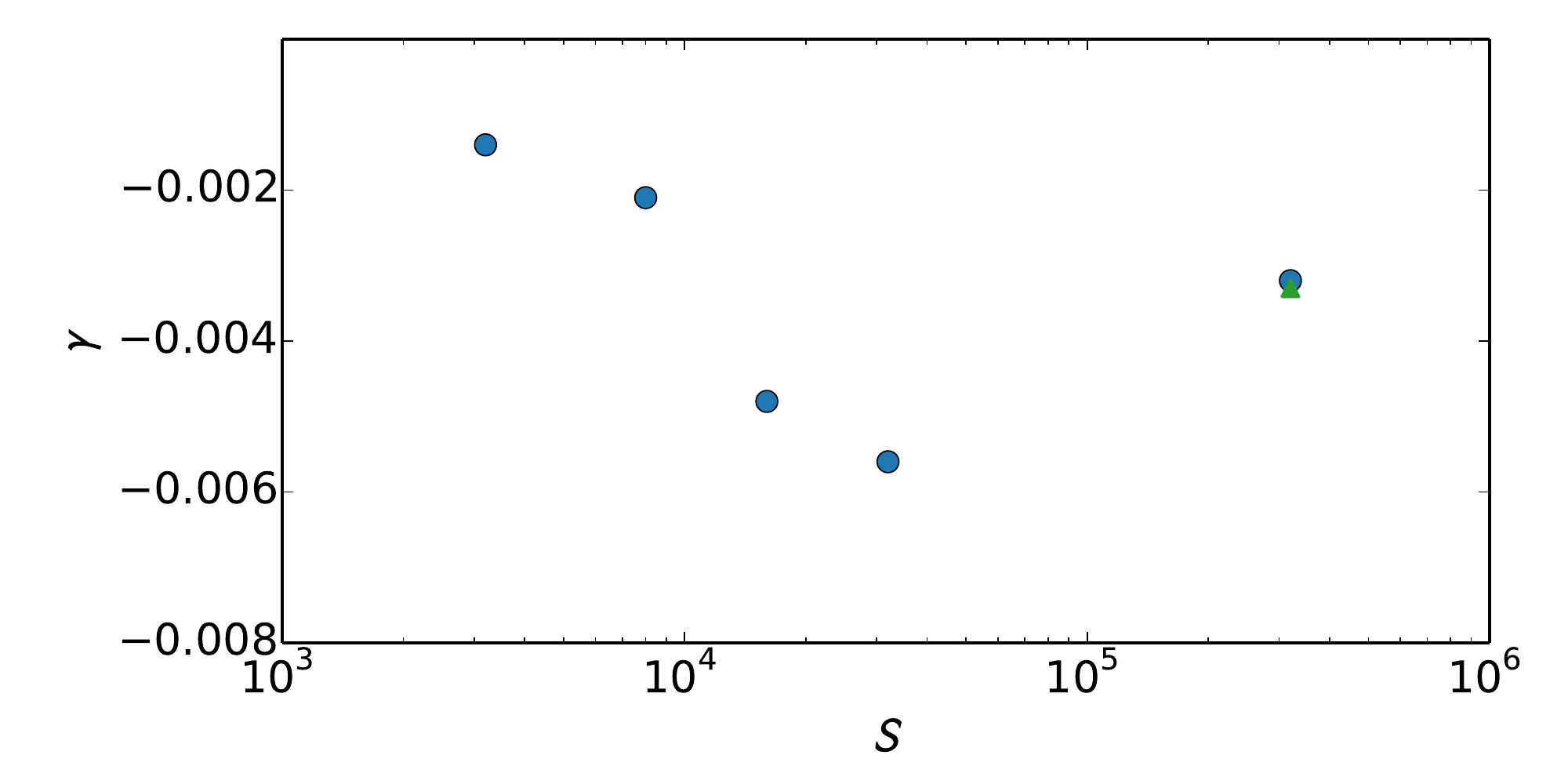}
  \caption{Decay rate $\gamma$ of the magnetic energy as a function of
    Lundquist number $\Lund$. Over two orders of magnitude in $\Lund$,
    $\gamma$ varies by less than a factor of three,
    non-monotonically. The triangle almost superposed on the highest
    $\Lund$ model shows the result for our double resolution run A*,
    demonstrating the excellent convergence we have achieved.}
  \label{fig:gamma_vs_s}
\end{figure}

Figures~\ref{fig:energy} and \ref{fig:current} show three phases of
reconnection: the slow, Sweet-Parker phase while linear instabilities
grow, a rapid exponential phase in which 3D effects dominate
reconnection, and finally a saturated, MHD turbulent phase. The
Sweet-Parker reconnection rate is $\propto \Lund^{-1/2}$, 
%mm leading to
     implying far slower than observed 
reconnection at the large values of $\Lund$ typical of Solar and
space plasmas. Figure~\ref{fig:gamma_vs_s} shows the decay rate
$\gamma$ during the rapid reconnection phase as a function of
$\Lund$. While $\Lund$ runs over two orders of magnitude, $\gamma$
varies by a factor of only about three, with no discernible functional
relationship to $\Lund$. Thus, the 3D instability offers a fast
reconnection mechanism that occurs at a rate apparently independent of $\Lund$,
without appeal to either kinetic effects or anomalous resistivity.

Once the instability saturates, the current sheet thickens
considerably (see the right panel of Fig.~\ref{fig:current}), and
the picture of a steady flow of fresh field from upstream (i.e. from
the $y$ direction above and below the layer) no longer holds in our
simulations. At the end of our simulation, there is still plenty of
field left to reconnect. The turbulence self-consistently driven in
the reconnection layer allows stochastic reconnection to occur
\citep{Lazarian1999,Eyink2011,Eyink2013}. The thickening of the
reconnection layer we see is consistent with their model: the
diffusion of the large-scale magnetic fields controls the ultimate
reconnection rate. Our periodic boundary conditions, similar to those
of \citet{2013arXiv1301.7424B}, do not allow for a self-consistent
steady state to occur. However, during the rapid,
resistivity-independent phase of the instability, the reconnection region grows.

\section{Discussion}
\label{sec:discussion}
We have shown that for Lundquist numbers $\Lund > 3.2 \times 10^{3} $,
current sheets become turbulent in the direction perpendicular to the
field along the sheet, leading to fast, 3D, magnetic reconnection
(i.e. decay rate of magnetic energy independent of resistivity
$\eta$). At high $\Lund$, individual 2D plasmoids rapidly lose their
identities, as the current sheet splits into filaments parallel to the
field direction.  These 3D effects, driven by rapidly growing
instabilities along current sheets, appear essential to understanding
reconnection. This provides strong support to the geometric ideas
advanced by
\citet{2012PhPl...19k2901B,2012PhPl...19i2902B,2013PhPl...20c2903B} as
well as the turbulent reconnection model developed by Lazarian,
Vishniac, and coworkers
\citep{Lazarian1999,2015ASSL..407..311L,2015arXiv150201396L}. To
demonstrate the latter point, in the
bottom panel of Figure~\ref{fig:energy} we show the reconnection
layer width $\Delta$ as a function of time. During the rapid growth
phase, $\Delta \propto t$, in pleasing agreement with the theory
described in \citet{2015arXiv150201396L}. We find that the layer expansion
velocity is $d\Delta/dt \sim 3 \times 10^{-3} v_A$, roughly a factor
of 
%mm  four  [there is some variation in Figure 3 of B13]
4--5 smaller than that reported by \citet{2013arXiv1301.7424B}. We
suspect the discrepancy is due to the fact that his simulations
include a guide field, although the lower diffusivity of his
pseudo-spectral code could also play a role. Nevertheless, the
agreement on the linear form of the growth rate despite our different setups and
codes supports the turbulent reconnection model.

As a result of the 3D instability, the initial current sheet develops
into a thick region of MHD turbulence. \citet{2015arXiv150201396L}
speculate that the growth in the plane perpendicular to the
reconnection (i.e. along the $z$ direction in our simulations) could
be due to Kelvin-Helmholz instability. Once the turbulent state is
reached, the decay of magnetic energy in our model slows dramatically,
back to a rate comparable to the initial Sweet-Parker rate. However,
we stress that the state of the system after the rapidly reconnecting,
unstable phase is radically different from the state before it, and
the slow subsequent reconnection may depend on the geometry we chose
for our simulations, which does not continue to force the system on
large scales, unlike, for example Solar flares
\cite{2014ApJ...784..144D}.  However, this turbulence naturally
produces the conditions required for fast stochastic reconnection
\citep{Lazarian1999,Eyink2011,Eyink2013}.

Future work employing deeper grid hierarchies, adaptive resolution
elements placed in regions of high $\mathbf{J}$, and simulations with
large scale forcing through inflow boundary conditions will clarify
the outcomes of the instability and its role in understanding
reconnection in astrophysical environments. Specifically, by removing
periodic boundary conditions (or isolating them via deep AMR
hierarchies), we will be able to make a more detailed study of a key
prediction of the \citet{Lazarian1999} model, the rate of broadening of
reconnection layer. 

\section{Acknowledgments}
\label{sec:ack}
JSO \& M-MML acknowledge support from NSF grant AST10-09802. MT
participated in the REU program at AMNH, funded by NSF grant
AST10-04591. We thank the anonymous referees for comments that
significantly improved the paper. We also thank A. Boozer,
C. R. DeVore, and A. Hubbard for useful discussions. This work used
the Extreme Science and Engineering Discovery Environment (XSEDE),
which is supported by National Science Foundation grant number
OCI-1053575. The computations presented here were performed on
\emph{Kraken} at the National Institute for Computational Science,
under grant TG-AST120045, and \emph{Stampede} at the Texas Advanced
Computing Center under grants TG-MCA99S024 and TG-AST140008.

\end{document}